\documentclass{Interspeech}
\usepackage{xcolor}
\usepackage{listings}

\usepackage{mdframed}

\usepackage{listings}
\usepackage{xcolor}

\definecolor{mintedbackground}{rgb}{0.95,0.95,0.96}
\mdfdefinestyle{mintedframe}{
    backgroundcolor=mintedbackground,
    linecolor=black,
    linewidth=0.5pt,
    innerleftmargin=2mm,
    innerrightmargin=2mm,
    innertopmargin=2mm,
    innerbottommargin=2mm,
    skipabove=0pt,
    skipbelow=0pt
}
\lstdefinestyle{mintedstyle}{
    backgroundcolor=\color{mintedbackground},
    basicstyle=\footnotesize\ttfamily,
    numbers=left,
    numberstyle=\tiny\color{black},
    breaklines=true,
    breakatwhitespace=true,
    tabsize=2,
    frame=single,
    framesep=2mm,
    xleftmargin=2mm,
    xrightmargin=2mm,
    showstringspaces=false,
    columns=flexible,
    keepspaces=true,
    numbersep=5pt
}

\interspeechcameraready

\title{In-context learning capabilities of Large Language Models to detect suicide risk among adolescents from speech transcripts}

\author[affiliation={1}]{Filomene}{Roquefort}
\author[affiliation={1}]{Alexandre}{Ducorroy}
\author[affiliation={1}]{Rachid}{Riad}

\affiliation{}{Callyope}{France}
\email{rachid@callyope.com}

\usepackage{comment}

\begin{document}

\maketitle

\begin{abstract}
    Early suicide risk detection in adolescents is critical yet hindered by scalability challenges of current assessments. This paper presents our approach to the first SpeechWellness Challenge (SW1), which aims to assess suicide risk in Chinese adolescents through speech analysis. Due to speech anonymization constraints, we focused on linguistic features, leveraging Large Language Models (LLMs) for transcript-based classification. Using DSPy for systematic prompt engineering, we developed a robust in-context learning approach that outperformed traditional fine-tuning on both linguistic and acoustic markers. Our systems achieved third and fourth places among 180+ submissions, with 0.68 accuracy (F1=0.7) using only transcripts. Ablation analyses showed that increasing prompt example improved performance ($p$=0.003), with varying effects across model types and sizes. These findings advance automated suicide risk assessment and demonstrate LLMs' value in mental health applications.
\end{abstract}

\section{Introduction}

Suicide represents one of the most pressing global public health challenges, ranking as the fourth leading cause of death among adolescents aged 15-19 years \cite{ward2021global}. The COVID-19 pandemic has further intensified this crisis, with studies reporting significant increases in suicidal ideation among young people \cite{bersia2022suicide}. Despite substantial efforts in suicide prevention, current assessment methods face critical limitations: psychological self-reports and professional evaluations often fail to capture the full extent of suicidal ideation and attempts, as individuals may be reluctant to disclose completely their thoughts before transitioning to a suicide attempt \cite{nock2008cross}.

The early detection of suicide risk is paramount for effective intervention and resource allocation. However, traditional assessment approaches, such as clinical interviews and psychological evaluations, face significant scalability challenges due to the global shortage of mental health professionals \cite{bishop2016population}. This scarcity is particularly acute in low- and middle-income countries \cite{scheffler2011human}, where the ratio of mental health workers to population can be as low as 1 per 100,000 people \cite{kakuma2011human}. These constraints underscore the urgent need for objective, scalable, and automated methods to detect suicide risk efficiently and reliably.

\begin{figure*}[!ht]
    \centering
    \includegraphics[width=1\linewidth]{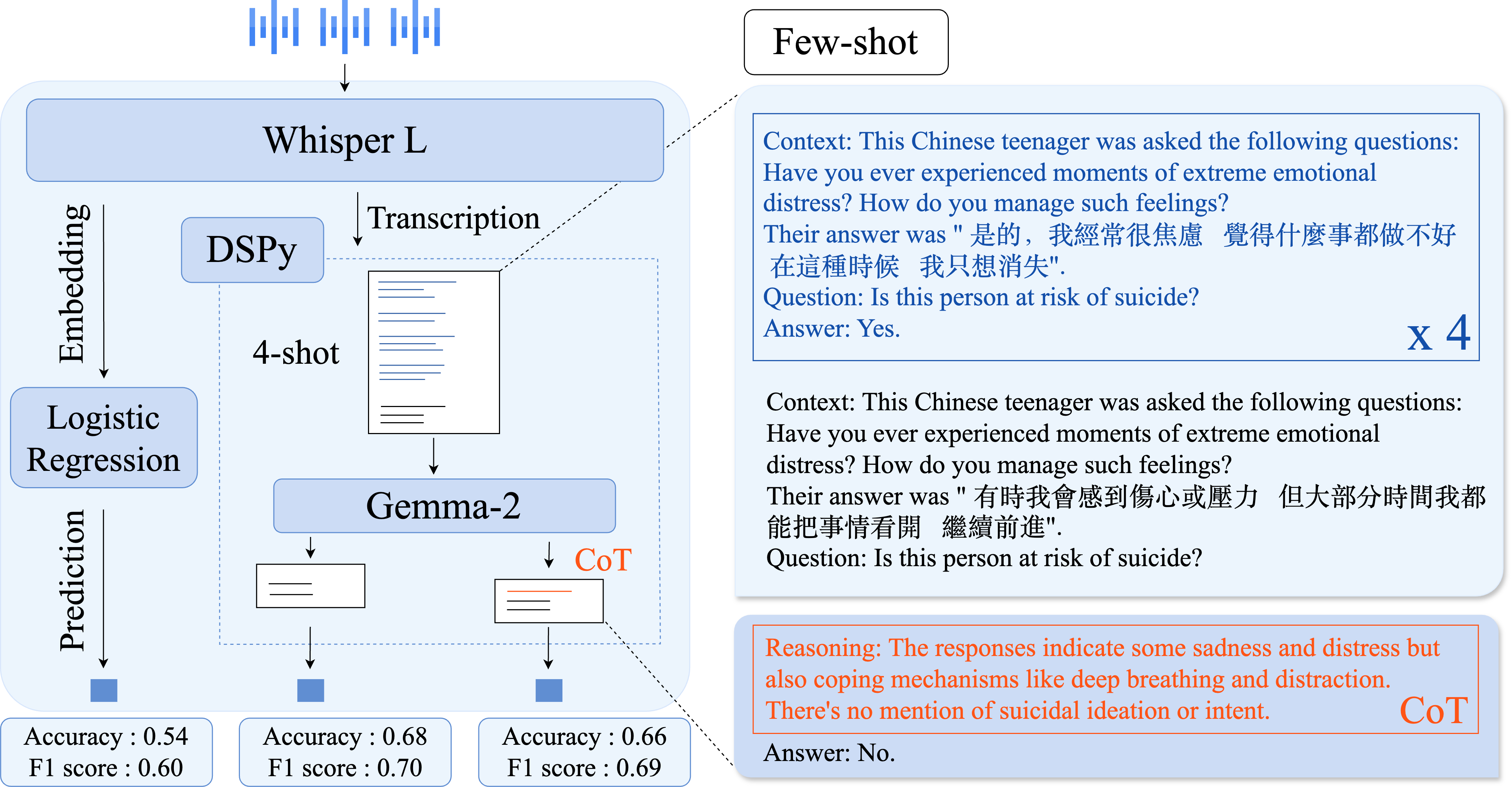}
    \caption{(Left) Overview of the three methods used for our submissions to the Speech Wellness Challenge. The first method employs direct audio processing using a pretrained audio encoder paired with a classifier. The second method uses a Large Language Model with a 4-shot classification approach. The third method extends this approach by adding chain-of-thought instruction. (Right) Illustration of the few-shot prompting technique and the chain-of-thought reasoning. The transcriptions are fake for privacy reasons, but the reasoning path is extracted from a successful prediction of the LLM on the dev set.}
    \label{fig:main_figure}
\end{figure*}

Speech markers have emerged as a promising avenue for identifying various mental health conditions \cite{cummins2015review}, including depression \cite{williamson2013vocal}, anxiety \cite{teferra2022screening}, and suicidal ideation \cite{hashim2012analysis, ding2025speech}. Both acoustic \cite{scherer2013investigating, stasak2021read} and linguistic representations \cite{homan2022linguistic} have been explored for speech-based suicide risk assessments. However, the widespread adoption of these automatic assessments in clinical practice has been hindered by several factors: the scarcity of publicly available datasets to develop robust methods, the absence of standardized benchmarks for system evaluation, and challenges in comparing different techniques \cite{wu20251st}.

The first SpeechWellness Challenge (SW1) \cite{wu20251st} was established to address these limitations by advancing speech techniques for detecting suicide risk through a mutual open benchmark. The challenge provides a unique dataset of 600 Chinese adolescents aged 10-18 years, with balanced representation of individuals identified as having suicide risk based on psychological scales. Participants were tasked with developing models that utilize both spontaneous and reading speech as digital biomarkers for binary suicide risk classification.
In this paper, we present our approach to the SW1 challenge, focusing primarily on Large Language Models (LLMs). Even though there are identified risk factors such as rumination correlated linked to suicide risk\cite{rogers2017rumination}, there is no clear and established way how to capture it automatically from speech content. Our additional decision to concentrate on linguistic features was motivated by our empirical findings that acoustic methods showed poor generalization, likely due to the challenge's speech anonymization requirements. These requirements particularly impacted speaker representations like deep speaker embeddings, which have previously shown promise in mental health and suicide risk assessment \cite{cui24_interspeech, gerczuk24_interspeech}. By leveraging only speech transcripts, our systems achieved third and fourth places among over 180 submissions, with an accuracy of 0.68, and surpassing the classic and challenging baselines from the challenge organizers.

Central to our methodology is the application of Large Language Models (LLMs) as few-shot learners \cite{brown2020language}. Rather than relying on manual prompt engineering, we employed a 'programming' approach to employ LLMs and their configuring by using the DSPy framework \cite{khattab2024dspy} to systematically develop and optimize our prompting strategy. This automated approach not only enhanced the robustness of our experiments but also ensured their reproducibility (a crucial consideration for clinical applications). Besides, we conducted extensive analysis and statistical experiments across multiple LLM architectures, number of in-context examples and model sizes to examine main contributing factors for suicide risk detection. Finally, we investigated Chain-of-Thought (CoT) reasoning approaches, performing qualitative analyses of the reasoning traces to understand how models interpret and process speech transcripts for suicide risk assessment.

Previous research by \cite{hashim2012analysis, ding2025speech, gerczuk24_interspeech} has greatly advanced speech-based approaches for suicide risk assessment, yet there remains no clear consensus on optimal detection methods from speech content across languages, particularly for adolescent populations. Besides, the reliability of acoustic markers can be compromised by recording devices \cite{jannetts2019assessing} and environmental factors like reverberation \cite{dineley23_interspeech}.

On the other hand, automatic speech recognition has made great strides in terms of performance, enabling reliable use of speech content \cite{radford2023robust}. Simultaneously, the progress of LLMs in numerous linguistic tasks opens new opportunities in mental health applications. \cite{schulhoff2024prompt} and \cite{skianis-etal-2024-leveraging} demonstrated that programmatic prompt engineering with DSPy outperformed manual prompt optimization for detecting mental health issues in social media content. Similarly, LLM applications have proven effective for PTSD detection \cite{quillivic2024interdisciplinary} and identifying OCD patterns \cite{kim2024large}.
Particularly relevant to our work, \cite{cui24_interspeech} achieved strong performance using both audio and language modalities on a non-anonymized version of the SpeechWellness Challenge dataset. 

\section{Speech wellness challenge}

The SW1 challenge consists of a suicidal risk detection task, where participants are required to produce a model to predict the label (has suicidal risk or not) for a subject. The training and testing datasets comprise speech recordings from 600 Chinese teenagers aged 10-18 years, with 50\% identified as having suicidal risk based on psychological scales \cite{sheehan2010reliability}. All subjects (speakers) have been anonymized to make the dataset available to participants of the SW1 challenge. This provides a unique opportunity to apply and refine advanced speech technologies for public health while preserving the privacy of all subjects.

Each speaker completed three audio tasks: two spontaneous responses and one reading task. The first spontaneous task required answering the question: "Have you ever experienced moments of extreme emotional distress? How do you manage such feelings?". The second open-ended task involved describing an image of a face expressing negative emotions. The reading task consisted of reading a passage from Aesop’s Fables untitled \textit{The North Wind and The Sun}. For our first approach using audio processing, we used all three audio tasks. But when focusing on speech transcripts, we only used the spontaneous tasks.

The speech recordings were preprocessed by the challenge organizers with neural voice conversion techniques to alter the voice's timbre and to anonymize the speakers with x-vector.
The dataset is split in two classes according to a binary label "at risk" or "not at risk". We reported the accuracy on the Dev set, and also the ranks of the participants and accuracy on the testing set for our 3 submissions. For more analysis, the challenge organizers provided the confusion matrix for our best system. 

\begin{table}[!ht]

  \centering
    \caption{Characteristics of the SW1 dataset for train, development (dev) and test sets.}
  \begin{tabular}{lllll}
    \toprule
    \textbf{}                & \textbf{All data}    & \textbf{Train}   & \textbf{Dev}      & \textbf{Test}  \\
    \midrule
    N        & 600        & 400       & 100        &100                \\
    Sex(F/M)                & 420/180                &  280/120       &  66/34         &   74/26    \\
    Age             & 13.8 (2.4)          & 13.8 (2.4)       & 13.8 (2.4)        &  13.8 (2.5)  \\
    Suic. risk           & 300/300              & 200/200          & 50/50             & 50/50          \\

    \bottomrule
  \end{tabular}

  \label{tab:word_styles}
  \vspace{-2.em}
\end{table}




\section{Methods}
\subsection{Baselines}

First, we tested if the audio modality is capable of detecting suicide risk. We compared classic signal processing, eGeMAPS features \cite{eyben2010opensmile}, and pretrained speech foundation models Whisper \cite{radford2023robust} and Hubert \cite{hsu2021hubert}. These speech foundation models are repurposed to provide a fixed-size embedding and provided to a classifier (Left part in Figure \ref{fig:main_figure}). The three audio tasks were segmented into 10-second slices. These slices were processed by each pretrained speech foundation model and labelled with the suicide risk label. A trained Logistic Regression classifier generated a prediction probability for each audio slice. We experimented with using each vocal task separately as well as in combination, applying different pooling techniques (mean, max and various mellowmax poolings \cite{asadi2017alternative}). 

As we observed poor generalization of participants on the global ranking, to ensure the capability of generalization of our methods, we reshuffled the training and development sets using stratified splitting. The folds remain balanced and with a similar distribution of age and sex. 
We reported the two baselines provided by the challenge organizers: one classic (1) using eGeMAPS followed by a Support Vector Machine (SVM) classifier to perform the classification task, and a (2) multimodal approach combining wav2vec 2.0 as audio encoder and BERT-BaseChinese model as text encoder (see \cite{wu20251st} for more details). For the ablation study (Table \ref{tab:ablation_studies}), we ran our experiments with different seeds for example selection and LLM inference.

\subsection{LLM as classification module}


Our approach leveraged LLMs as binary suicide risk classifiers from speech transcripts obtained with Whisper \cite{radford2023robust} through in-context learning. Following \cite{radford2019language, brown2020language}, we prompted each LLM to perform classification without any fine-tuning, by providing task descriptions and examples within the context window. We explored both standard zero-short inference \cite{radford2019language}, few-short inference \cite{brown2020language} and chain-of-Thought (CoT) reasoning \cite{wei2022chain} as illustrated in Figure \ref{fig:main_figure}. The main hypothesis is that examples can help LLMs find patterns and improve probability distribution for language modeling. The main idead behind CoT, it encourages models to make intermediate reasoning steps before making a final prediction. 

The classification pipeline was developed with the DSPy framework \cite{khattab2024dspy}, which enabled programmatic prompt construction, ensured experimental reproducibility and allowed easy inspection of outputs. Our prompt template that structured the input with specific contextual framing is displayed in Figure \ref{fig:prompt_template}.

   



\begin{figure}[!ht]
\begin{mdframed}[style=mintedframe]
\begin{lstlisting}[numbers=left, numberstyle=\tiny, basicstyle=\footnotesize\ttfamily, frame=none]
 [[ ## context ## ]]
 A Chinese teenager was given 2 tasks.
 1. They had to answer to the 
 following question 
 'Have you ever experienced moments of
 extreme emotional distress? 
 How do you manage such feelings?'
 Their answer: {first speech transcript}
 2. They were shown an image of a face 
 displaying negative emotions, and asked
 to describe it.
 Their answer: {second speech transcript}
 
 [[ ## question ## ]]
 Is this patient at risk of suicide?
\end{lstlisting}
\end{mdframed}
\centering
\vspace{-1.em}
\caption{Prompt template for our submission to define the DSPY program to tackle suicide risk detection based on speech transcripts.}
\label{fig:prompt_template}
\vspace{-1.em}
\end{figure}
We compared two LLM models that showed great performance in multiple benchmarks: Gemma2 \cite{team2024gemma} with 9 billions of parameters and Qwen2.5 with 7 billions of parameters \cite{yang2024qwen2}. We chose the models versions that were fine-tuned with instructions. For ablation analysis, we also studied a smaller and larger version of Gemma2 with 2 and 27 billions of parameters.

\subsubsection{Statistical analysis of in-context learning}
After the SW1 challenge, to analyze the relationship between classification accuracy and the number of examples provided in the LLM prompt, we performed a large scale experiments with multiple models version and size. We computed the accuracy as a function of the number of examples. We ran a multiple linear regression with interaction terms. The model was specified as:

\begin{align*}
\text{Accuracy} &\sim \text{example\_count} + C(\text{model\_type}) + \text{model\_size} \nonumber \\
&\quad + \text{example\_count} \times C(\text{model\_type}) \\ &\quad+ \text{example\_count} \times \text{model\_size}
\end{align*} 

Here are the definitions of each term: $\text{example\_count}$ represents the number of examples included in the prompt. $C(\text{model\_type})$ is a categorical variable representing different LLM models, and $\text{model\_size}$ denotes the size of the model.  
We also included interaction terms to examine whether the effect of increasing the number of examples varies across different models. Specifically, $\text{example\_count}:C(\text{model\_type})$ captures the interaction between the number of examples and model type, while $\text{example\_count}:\text{model\_size}$ examines how the effect of additional examples changes with model size.

We used Ordinary Least Squares (OLS) regression to estimate coefficients and test their significance.
We ran 3 times the experiments with different seeds to ensure reliability of our ablation analysis, and reported each coefficient ($\beta$) and its associated $p-$value.

\section{Results}


Our experiments with classic acoustic features showed limited success (See Table \ref{tab:submitted_results}). The baseline OpenSMILE features \cite{eyben2010opensmile} performed poorly, which our replication confirmed. This underperformance likely stems from the anonymization process applied to the audio data. We achieved optimal acoustic model performance by applying a mellowmax(1.0) pooling to slice-level predictions using Whisper L features. This approach yielded an accuracy of 0.63 on the Dev* set and 0.54 on the Test set. Compared to the original SW1 authors' work \cite{cui24_interspeech}, the audio modality seem compromised by anonymization.

Zero-shot approaches with LLMs showed modest results, with both Qwen2.5 and Gemma2-9b achieving only 0.52 accuracy on the Dev set.
However, in-context learning demonstrated promising results both for Qwen2.5 and Gemma2, with a potential positive link between performance and the number of examples included in the prompt. Gemma2-9b particularly stood out with superior performance compared to other models. Our best-performing submission leveraged Gemma2-9b with 4-shot prompting, achieving 0.67 accuracy on the development set and 0.68 on the test set, securing 3rd place in the challenge. Finally, Chain-of-Thought (CoT) reasoning did not significantly improve performance, yet still achieved 4th place overall.

\begin{figure}[!ht]
    \centering
    \includegraphics[width=\linewidth]{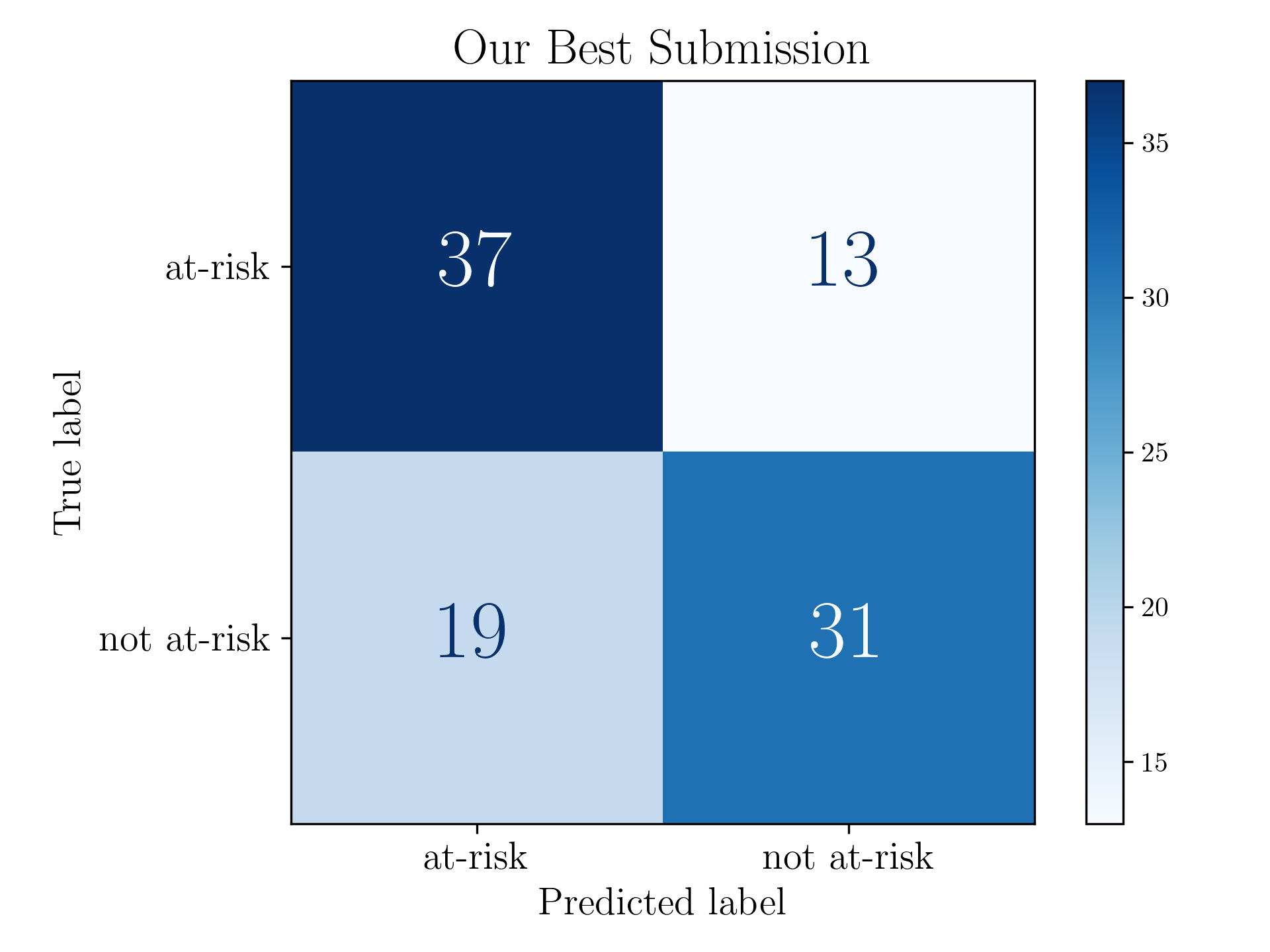}
    \caption{Confusion matrix for our best system on the test set.}
    \label{fig:confusion_matrix}
    \vspace{-1.em}
\end{figure}

We also reported the confusion errors on the test set for our best system (Figure \ref{fig:confusion_matrix}) and found that our method prioritizes detecting at-risk individuals. This bias reduces false negatives — critical in suicide risk detection — while increasing false positives, a less harmful trade-off in this situation. 

In Table \ref{tab:ablation_studies}, we reported the results of few-shot classification across different model architectures and sizes. When performing few-shot classification, LLMs selects examples from the training set —we used random sampling— and applies the same examples for all predictions on the development set. To evaluate the impact of examples selection, we varied DSPy's random seed and observed performance variation due to example selection. For instance, our submission configuration (4-shot with Gemma-2-9b) shows now a standard deviation of 0.7 in accuracy when evaluated with three different seeds, while default seed got 0.67 accuracy.  This variability highlights the significant influence of example selection on model performance. Example selection in the prompt seem to play therefore a critical part for generalization as reported by the authors of DSPy framework on other benchmarks \cite{khattab2024dspy}.

\begin{table}[!ht]
    \centering
        \caption{Final Accuracy results for the SW1 challenge on the different sets. Not all approaches could be evaluated on the held out Test set during the SW1 challenge period. LLMs experiments realized with DSPy on a single seed.}
    \begin{tabular}{lllll}
        & Dev* & Test & Rank \\
        \hline
        Baselines \\
        \hline
        eGeMAPS + SVM \cite{cui24_interspeech} & / & 0.51 & 130/188 \\
        Wav2Vec 2.0 + BERT \cite{wu20251st} & / & 0.61 & 15/188 \\
        \hline
        Audio representations \\
        \hline
        eGeMAPS + Logistic Regression & 0.51 & / & / \\
        Hubert L + Logistic Regression & 0.53 & / & / \\
        Whisper L + Logistic Regression& 0.63 & 0.54 & 82/188 \\
        \hline
        LLM approaches \\
        \hline
        Qwen2.5-7b Zero-shot & 0.52 & / &  / \\   
        Qwen2.5-7b 1-shot & 0.55 & / &  / \\ 
        Qwen2.5-7b 4-shot & 0.59 & / &  / \\
        Gemma2-9b Zero-shot & 0.52 & / &  / \\   
        Gemma2-9b 1-shot & 0.55 & / &  / \\ 
        Gemma2-9b 4-shot & \textbf{0.67} & \textbf{0.68} &  3/188 \\ 
        Gemma2-9b 4shot — CoT & \textbf{0.67} & 0.66 &  4/188 \\   
        \hline
        \hline
        First place of the challenge & / & \textbf{0.74} & 1/188   \\  
        \end{tabular}
\vspace{-1.5em}
    \label{tab:submitted_results}
\end{table}

\begin{table}[!ht]
    \centering
    \small
        \caption{Results of classification experiments with varying model architectures and sizes (Gemma-2 2b/9b/27b Instruct, Qwen2.5 7b Instruct), and few-shot settings, k being the number of examples provided to the model. Mean Accuracy and standard deviation are computed over three runs with different random seeds. 64 and 128 examples could not compute for Gemma2 2b and 9b.}
    \setlength{\tabcolsep}{2pt} 
    \begin{tabular}{lccccc}
        \hline
        k & \textbf{Gem.2b} & \textbf{Gem.9b} & \textbf{Gem.27b} & \textbf{Qwen7b} & Avg.\\
        \hline
        0  & 0.52 (.00) & 0.52 (.00) & 0.53 (.00) & 0.52 (.00) & 0.52 (.00)\\
        1  & 0.55 (.01) & 0.57 (.04) & 0.57 (.03) & 0.55 (.02) & 0.56 (.01)\\
        2  & 0.52 (.02) & 0.57 (.05) & 0.56 (.01) & 0.57 (.03) & 0.55 (.02)\\
        4  & 0.54 (.06) & 0.60 (.07) & 0.58 (.02) & 0.60 (.05) & 0.58 (.02)\\
        8  & 0.53 (.03) & 0.60 (.06) & 0.58 (.02) & 0.61 (.03) & 0.58 (.03)\\
        16 & 0.57 (.03) & 0.59 (.04) & 0.61 (.04) & 0.59 (.01) & 0.59 (.01)\\
        32 & 0.57 (.04) & 0.64 (.01) & 0.61 (.04) & 0.58 (.03) & 0.60 (.03)\\
        64 & N/A & N/A & 0.62 (.00) & 0.57 (.05) & 0.60 (.03)\\
        128 & N/A & N/A & 0.53 (.03) & 0.56 (.04) & 0.55 (.02)\\
        \hline
        Avg. & 0.54 (.02) & 0.58 (.03) & 0.58 (.03) & 0.57 (.03) \\
        \hline
    \end{tabular}

    \vspace{-1.em}
    \label{tab:ablation_studies}
\end{table}

The results from the OLS regression indicate that increasing the number of examples in the prompt significantly improves classification accuracy ($\text{example\_count}$ $\beta= 0.0023$, $p = 0.003$), supporting the hypothesis that in-context learning enhances model performance. However, the interaction terms reveal that this effect is not uniform across models. The Qwen model shows a smaller benefit from additional examples compared to the baseline, as indicated by the negative interaction term ($\text{example\_count}$:C($\text{model\_type}$)[T.qwen] $\beta =-0.0017$, $p = 0.005$), suggesting that Qwen may rely less on in-context learning. Additionally, larger models tend to achieve higher accuracy ($\text{model\_size}$ $ \beta= 0.0018$, $p = 0.003$), but their benefit from additional examples is reduced, as shown by the negative interaction between example count and model size ($\text{example\_count}$:$\text{model\_size}$ $\beta= -8.7e-05$, $p = 0.002$). However, the effect size for this interaction is very small, suggesting that while significant, this interaction may not have a meaningful impact in practical settings. The overall model explained a modest portion of the variance ($R^2=0.134$), this suggests that these factors influence accuracy but there are other factors explaining the full accuracy. 


The findings presented in this study are based on the scoring framework of the MINI-KID scale \cite{sheehan2010reliability}, which assesses current suicide risk as at risk or no risk. This classification reflects participants’ immediate responses to the MINI-KID assessment and should not be interpreted as a prediction of future suicidal behavior. Although the MINI-KID suicide module is widely recognized as a gold standard for assessing current suicide risk among adolescents, it has limitations. It relies heavily on self-reported data, which may lead to underreporting or misinterpretation of symptoms, and its fixed set of items may not fully capture the complex and dynamic nature of suicidal ideation and behavior. Accordingly, the results reported herein are strictly confined to the context of this assessment. 
\section{Conclusions}

Our work on the SW1 demonstrated the effectiveness of LLM-based approaches for suicide risk detection despite speech anonymization constraints. By leveraging in-context learning with DSPy, our system achieved competitive results using only speech transcripts data, securing third place in the challenge. Statistical analysis confirmed that increasing example count significantly improves classification accuracy, though this effect varies across model types and sizes.
In future work, we aim to examine CoT reasoning paths more deeply to extract linguistic patterns used in accurate predictions. This analysis could provide valuable insights to mental health practitioners by revealing data-driven indicators of suicide risk in adolescents. By making these linguistic patterns interpretable to clinicians, our approach could bridge the gap between computational methods and practical clinical applications, potentially enhancing early intervention strategies for at-risk adolescents.
\newpage

\newpage

\bibliographystyle{IEEEtran}
\bibliography{mybib}

\end{document}